\documentclass[12pt,aps]{revtex4}
\usepackage{graphics}
\input epsf                  
\begin{document}
\draft
\title{Symmetry restoration of the soft pion corrections for the light sea
quark distributions in the small $x$ region}
\author{Susumu Koretune}
\address{Department of Physics,Shimane Medical University,Izumo,Shimane,693-8501,Japan}
\date{\today}          
                           
\begin{abstract}
The soft pion correction at high energy may play a crucial role in
non-perturbative parts of sea quark distributions. In this paper, 
we show that, while the soft pion correction for the strange sea qaurk 
distribution is suppressed in the large and the medium $x$ region
compared with that for the up and the down sea quark one, it can become 
large and $SU(3)$ flavor symmetric in the very small $x$ region. 
This gives us a good reason for the symmetry restoration of light 
sea quark distributions required by the mean charge sum rule for the 
light sea quarks. Then, by estimating this sum rule with the help of
the results obtained by the soft pion correction, 
it is argued that there is a large symmetry restoration
of the strange sea quark in the region from $x=10^{-2}$ to $10^{-6}$ 
at $Q^2\sim 1$ GeV.
 
\end{abstract}
\maketitle
\section{Introduction}
The soft pion theorem in exclusive reactions at low energy has been
well established. The same theorem can be applied to inclusive
reactions. However, in this case, what kind of physical processes could be
identified as the one in the soft pion limit was unclear. Many years ago,
an interesting proposal that the pions in the central region with the low
transverse momentum in the center of the mass (CM) frame might be
identified as the soft pion was given\cite{sakai}.
Though this proposal was turned out to be false, it had been recognized
that it might be physically meaningful if we restricted
the pions to the ones produced directly in the reaction, where the
directly produced meant that among the pions in the final state the ones in
decay products of resonance particles should be excluded\cite{kore78,kore78reso,kore82}.
In this sense, the proposal in Ref.\cite{sakai} had opened up the way to relate 
the soft pion theorem at high energy to physical reactions.  
Let us explain the fact in the semi-inclusive reactions 
$\pi (q) + N(p) \to \pi_{s}(k) + anythings(X_0)$, where $\pi_{s}$ is the soft pion,
and $anythings(X_0)$ includes no soft pion.
In the CM frame, we regard the directly produced pion
below a low transverse momentum and a small Feynman scaling
variable as the soft pion and the one above this cut as the hard
pion. Then we identify this soft pion as the one in the soft pion limit
through the refined scaling assumption which states that the
differential cross section of the directly produced pions 
divided by the total cross section behaves smoothly near $x_F=0$
for each energy. Here, the energy dependence of the value
of the normalized invariant cross section at $x_F=0$ is allowed. 
We call this refined scaling as the smoothness assumption.
Though the experimental value of the inclusive cross section
in general includes the multi-soft pion processes, by taking the
ratio with the total cross section, this multi-soft pion effect 
cancels out, and we can compare the theoretical value of the one 
soft pion process with the experimental value.
In this way, a theoretical ambiguous part in the
infra-red structure in the hadronic reaction originating from
the soft pion has been replaced by the experimental value,
and applicability of the soft pion theorem is extended to the high energy region.
Based on this observation, the soft pion
contribution to the Gottfried sum was investigated\cite{kore20}, 
and it was found that it gave a sizable contribution to it and that its magnitude 
was just the one to compensate the typical contribution based solely 
on the meson cloud model\cite{kumano}. This fact was consistent with 
the study based on the modified Gottfried sum
rule\cite{kore93,kore-ichep} in the sense that 
about $40 \%$ of the departure from the value $1/3$ came from the region 
where the momentum of the kaon in the laboratory frame was above 4GeV/$c$. 
In this paper, we derive the soft pion corrections for the light sea
quark distributions. 
In sect.II, we give a kinematics of the single soft pion observed
inclusive reaction. In sect.III, we give soft pion contribution
to the light sea quark distribution, and show that the soft pion correction
for the strange sea quark one is greatly suppressed compared
with that of the up and the down sea quark one in the large 
and the medium $x$ region. In sect.IV, we show that, under a certain condition, 
the soft pion correction for the light sea quark distributions becomes $SU(3)$ 
flavor symmetric in the very small $x$ region. Then, using the mean 
charge sum rule for the light sea quarks, we discuss the behavior of the light 
sea quark distributions in the small $x$ region.
In sect.V, we give a conclusion. In Appendix A, we give a detailed
explanation of the kinematics of the method,
and in Appendix B, we explain how the soft pion contribution to the 
phenomenologically determined up and down sea quarks enters.\\
\section{Kinematics}
Let us consider the semi-inclusive current induced reaction
$V_{a}^{\mu}(q) + N(p) \to \pi_{s}(k) + anythings(X_0)$, where $V_a^{\mu}$ is the
electromagnetic or weak hadronic currents,
and $anythings(X_0)$ includes no soft pion. The soft pion limit
of this reaction can be obtained by the Adler consistency
condition\cite{Adler}. By keeping the pion mass $m_{\pi}\neq 0$, we
take $k^{\mu}\to 0$ limit of the amplitude where the soft pion is
off-shell and the rest of the particles are on-shell.
We first take $k^{+}=0$ and $\vec{k}^{\bot}=0$, and after that
we take $k^{-}=0$. In this limit, $k^2$ is restricted to be 0, 
but the momentum of the initial particles are unrestricted. 
Here we use the PCAC relation
$\partial_{\mu}J_a^{5\mu}(x)=m_{\pi}^2F\phi_{\pi}(x)$, 
where $F=\sqrt{2}f_{\pi}$ for $a=1 \pm i2$ and $F=f_{\pi}$ for $a=3$.
The resulting expressions are given by the terms free from the pion pole terms
and the null-plane commutator term as in Fig.1.
\begin{figure}
   \centerline{\epsfbox{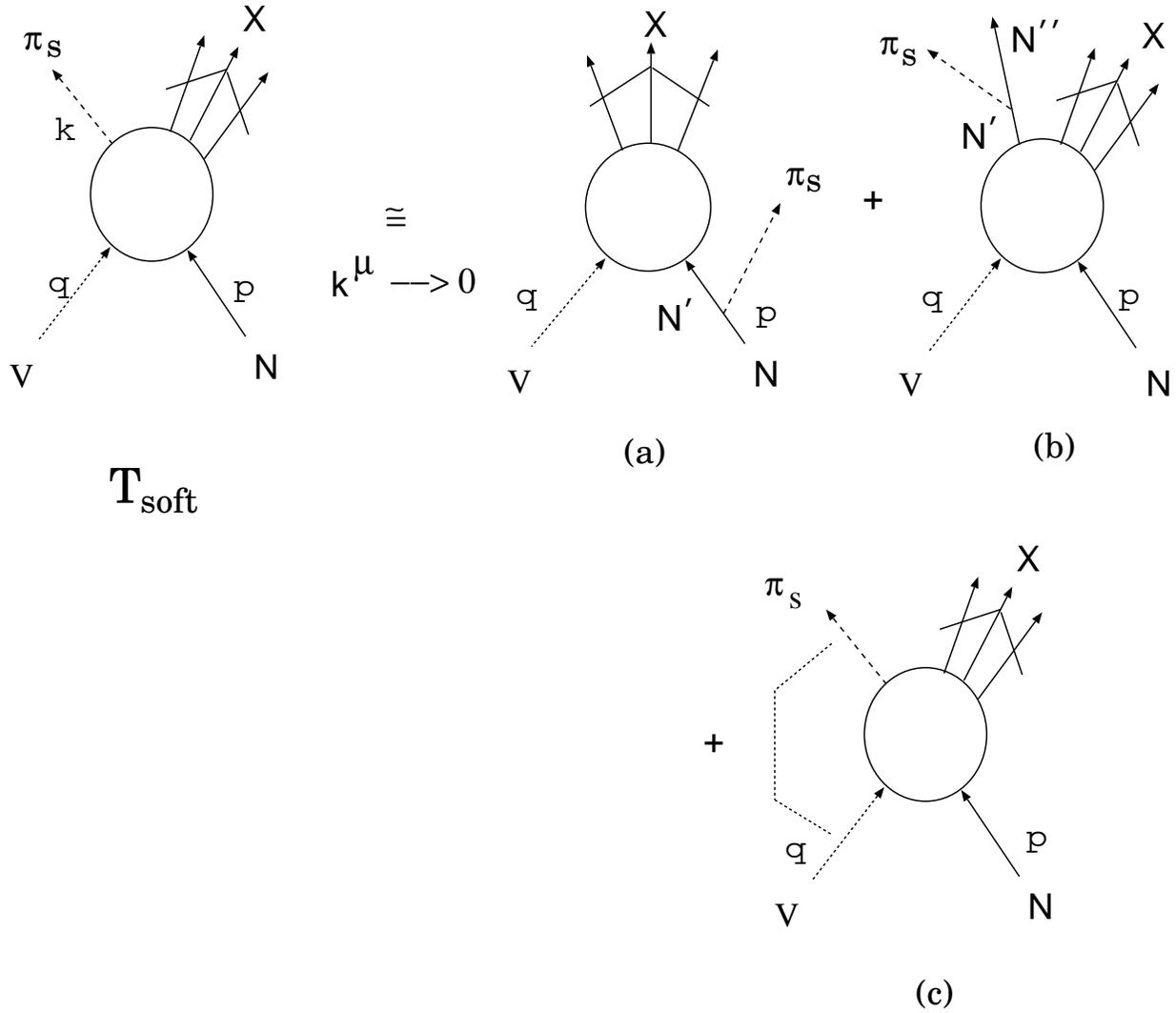}}
   \caption{The soft pion theorem in the inclusive reaction. The graph (a) 
is the one where the proper part of the axial-vector current is attached to the 
initial nucleon, the graph (b) is the one to the final nucleon
   (anti-nucleon), and the graph (c) is the one which comes from the null-plane commutator.}
        \label{fig:1}
\end{figure}
The graph (a) is the one where the proper part of 
the axial-vector current is attached to the initial nucleon. 
We call the term which comes from this type of the graph as 
the pion emission from the initial nucleon. 
The graph (b) is the one where the
proper part of the axial-vector current is attached to the final 
nucleon (anti-nucleon). We call the term which comes from
this type of the graph as the pion emission from the final nucleon. 
The graph (c) is the one which comes from the
null-plane commutator. We call the term of this kind as 
the commutator term.
The hadronic tensor can be obtained by squaring this amplitude,
and we have several terms corresponding to the three origins in the
amplitude. Among them, the term where the one soft pion is attached
to the one final nucleon (anti-nucleon) and the other to the initial
nucleon or the term where the two soft pions are attached to the
different final nucleons can be neglected at high energy. 
In these cases we have the odd number of the helicity factors 
in the same nucleon(anti-nucleon) line in the final state
arising from the matrix element of 
the following form 
$\langle p(p),h|J^{5+}_{a}(0)|n(p),h^{\prime}\rangle =
2p^+hg_A(0)\delta_{hh^{\prime}}$,
where $p$ means the proton, $n$ means the neutron,
$h$ means the helicity factor, and we take
$a=1 + i2$ by way of illustration.
Since, at high energy, we can expect that the production of 
the $+$ helicity nucleon (anti-nucleon) and that of the $-$ helicity one
are the same order, the contributions from these terms are expected to be small 
compared with those terms where 
the helicity factors in the same nucleon(anti-nucleon) line
in the final state are even.
Typical graphs contributing to the hadronic tensor are given in Fig.2.
\begin{figure}
   \centerline{\epsfbox{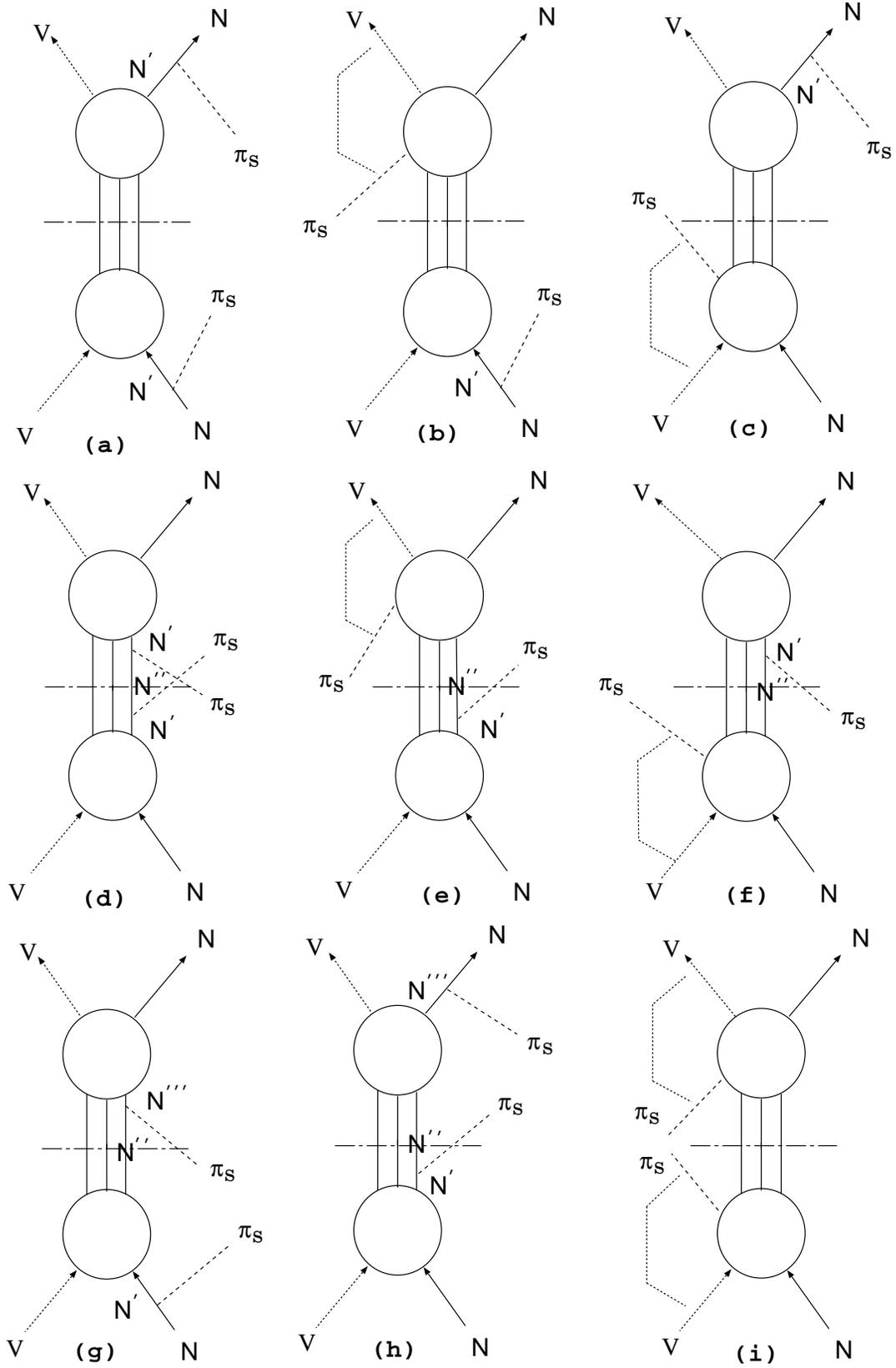}}
   \caption{Graphical representation of the soft pion correction for the hadronic tensor.}
        \label{fig:2}
\end{figure}\\
Thus, among the graphs in Fig.2, we take the contribution 
from the graphs $(a),(b),(c),(d)$, and $(i)$ into consideration,
and discard those from the graphs $(e),(f),(g)$, and $(h)$.
Now the contributions from the graphs $(a)$ and $(d)$ are related to
the known process directly. Hence, to estimate these,
we need no assumption except the
one necessary to apply the soft pion theorem. However, to estimate
the contribution from the graphs 
from $(b),(c)$,and $(i)$, we need further theoretical consideration.
A detailed explanation to estimate these parts are given in Appendix A.
\section{Soft pion contribution to the sea quarks}
Now we denote the structure functions $F_2^{lp(n)}$ 
where the suffix $l$ means the lepton and $p(n)$ means
the proton (the neutron). Further, we define
$F_2^{\mu N}= \frac{1}{2}(F_2^{\mu p} + F_2^{\mu n}),$ and 
$F_2^{\nu N}= \frac{1}{4}(F_2^{\nu p} + F_2^{\nu n} + F_2^{\bar{\nu} p} + F_2^{\bar{\nu} n} ).$
Let us first give the soft pion correction for the differential between
the up and the down sea quark distribution\cite{kore20}. Since the
separation of the sea and the valence part in the up and the
down quark distribution has some ambiguity, we use here the Adler sum
rule. This sum rule is considered to be exactly satisfied at the structure function
level. We require this also at the quark distribution level as is
usually fulfilled in a phenomenological analysis.
Thus the valence up and down quarks need no correction
from the soft pion since they are defined through the 
experimentally measurable quantity such as 
the structure function where the soft pion effect is already
included. This causes a subtlety in the separation
of the up and the down sea quark distribution into the soft pion part 
and the other one.
A detailed explanation of this fact is given in Appendix B.
Then, the soft pion
correction for $x(\lambda_d - \lambda_u)$ where $\lambda_i$ for $i=u,d,s$
denotes the sea quark distribution for each flavor
has been determined from the soft pion correction for the structure 
function $(F_2^{ep}-F_2^{en})$ as
$(F_2^{ep}-F_2^{en})|_{{\rm soft}}=-2x(\lambda_d - \lambda_u)|_{{\rm
soft}}/3$.
Here the $\lambda_d|_{{\rm soft}}$ and $\lambda_u|_{{\rm soft}}$
are $\lambda_d|_{{\rm soft}}^1$ and $\lambda_u|_{{\rm soft}}^1$
in Appendix B respectively(see Eqs.(B12) and (B13)). Thus we obtain
\begin{equation}
x(\lambda_d - \lambda_u)|_{{\rm soft}}
= \frac{-3I_{\pi}}{8f_{\pi}^2}[g_A^2(0)(F_2^{ep} - F_2^{en})(3<n> -1)
-16xg_A(0)(g_1^{ep} - g_1^{en})].
\end{equation}
The various expressions on the right-hand side of this equation are as follows: The mean
multiplicity $\langle n \rangle$ is the sum of the nucleon and the
anti-nucleon multiplicity, and the factor $I_{\pi}$ is the phase space factor of the
soft pion defined as
\begin{equation}
I_{\pi} = \int\frac{d^2\vec{k}^{\bot}dk^+}{(2\pi)^{3}2k^+},
\end{equation}
with a kinematical constraint explained later in this section.
The spin dependent structure function $g_1^{ep(n)}$ is
a usual one. For example, $g_1^{ep}$ can be expressed as
$\displaystyle{\frac{1}{2}\sum[\frac{4}{9}\triangle u +
\frac{1}{9}\triangle d + \frac{1}{9}\triangle s]}$, 
where $\triangle q$ with $q=u,d,s$ is a sum of the quark and the antiquark of the
differential between the helicity $+$ distribution and the helicity $-$ one along the
direction of the proton spin in the infinite momentum frame in the parton model. 
The spin-dependent term in Eq.(1) is obtained in the approximation in which
the sea quark contribution to $(g_1^{ep} - g_1^{en})$ is ignored. 
Without this approximation, $16(g_1^{ep} - g_1^{en})$ in Eq.~(1) should be replaced by
$\displaystyle{24(g_1^{ep} - g_1^{en})-\frac{4}{3}(g_1^{\bar{\nu} p} -
g_1^{\nu p})}$. The graphs in Fig.2 can be identified to the various
expressions in Eq.(1) as follows.
The term proportional to the $\langle n \rangle$
comes from the graph $(d)$, the term proportional 
to $g_A^2(0)$ without the nucleon multiplicity factor comes from the
graph $(a)$, and the term proportional to $g_A(0)$ comes 
from the graphs $(b)$ and $(c)$. 
The contribution from the graph $(i)$ is canceled out by
adding the contributions from the $\pi_s^+$,$\pi_s^-$, and $\pi_s^0$. 

The soft pion correction for the strange sea quark distribution
can be obtained by calculating the soft pion contribution to the
structure function $(\frac{5}{6}F_2^{\nu N} - 3F_2^{\mu N})$. 
In the $SU(4)$ model with the Cabibbo angle being 0 we obtain
\begin{eqnarray}
\lefteqn{(\frac{5}{6}F_2^{\nu N} - 3F_2^{\mu N})|_{{\rm soft}}=\frac{I_{\pi}}{f_{\pi}^2}[\frac{1}{12}(F_2^{\nu p}
+ F_2^{\bar{\nu} p})_0 + \frac{5}{16}g_A^2(0)(F_2^{\nu p}+ F_2^{\bar{\nu} p})(1+\langle n \rangle^{\nu N})}\nonumber\\
&-&\frac{9}{8}(F_2^{ep} + F_2^{en})g_A^2(0)(1+\langle n \rangle^{eN}) - 10xg_A(0)(g_1^{ep} - g_1^{en})].\hspace{4cm}
\end{eqnarray}
Since $(\frac{5}{6}F_2^{\nu N} - 3F_2^{\mu N})$ is expressed as 
$x(\lambda_s-\lambda_c)$, in the kinematical region
where the charm sea quark can be neglected, we can set
$(\frac{5}{6}F_2^{\nu N} - 3F_2^{\mu N})=x\lambda_s$.
The suffix 0 in the expression $(F_2^{\nu p}+ F_2^{\bar{\nu} p})_0$ 
on the right hand side of Eq.(3) means
the structure function defined in the $SU(3)$ model, 
and it comes from the graph $(i)$ in Fig.2. 
The reason why we meet here such structure functions
is as follows. The flavor suffix of the axial-vector current corresponding to
the pion is $1\pm i2$ or $3$. We use the commutation relation on the 
null-plane between the hadronic weak currents and the axial-vector current, 
hence the part related to the strange sea quark and the charm sea quark 
in the weak hadronic current drops out in this step. 
The structure function obtained after such a manipulation
is equivalent to the structure function in the
$SU(3)$ model with the Cabibbo angle being 0.
Thus in the kinematical region where the charm sea quark 
contribution is neglected we obtain
\begin{eqnarray}
x\lambda_s|_{{\rm soft}} = \frac{I_{\pi}}{f_{\pi}^2}[\frac{x}{6}(d_v + u_v)
+ \frac{x}{3}(\lambda_u + \lambda_d) + \frac{3}{4}xg_A^2(0)(1+\langle n \rangle)\lambda_s \\\nonumber
- \frac{5}{3}xg_A(0)(\triangle u_v - \triangle d_v)],
\end{eqnarray}
where we set $\langle n \rangle^{\nu N}=\langle n \rangle^{eN}=\langle n \rangle$, 
and the $\triangle u_v$($\triangle d_v$) is the valence part
in the  $\triangle u$($\triangle d$).
Now both hand sides of Eq.(3) get contribution from the charm sea
quark. The equation which does not neglect
the charm sea quark contribution is the one where the
$\lambda_s|_{{\rm soft}}$ on the left-hand side of
Eq.(4) is simply replaced
by the $(\lambda_s - \lambda_c)|_{{\rm soft}}$ and 
the $\lambda_s$ on the right-hand side
of it by the $(\lambda_s - \lambda_c)$. Since the main contribution 
in the small $x$ region comes from the term proportional to the 
factor $(1+\langle n \rangle)$ and the charm sea quark begins to
contribute also in this region, the additional parts which is
added to both hands side of Eq.(4) can be equated as
\begin{equation}
x\lambda_c|_{{\rm soft}} = \frac{I_{\pi}}{f_{\pi}^2}\left[\frac{3}{4}xg_A^2(0)(1+\langle n \rangle)\lambda_c \right].
\end{equation} 
With this assumption, we regard Eq.(4) as the formula to determine the soft pion contribution 
to the strange sea quark distribution even in the region where the charm sea quark contribution 
can not be neglected.
Now, following Eqs.(B12) and (B13) in Appendix B, we have the relation
$F_2^{\nu N}|_{{\rm soft}}=2x(\lambda_u|_{{\rm soft}} + \lambda_d|_{{\rm
soft}} + \lambda_s|_{{\rm soft}} + \lambda_c|_{{\rm soft}})$,
where $\lambda_u|_{{\rm soft}}=\lambda_u|_{{\rm soft}}^1$ and
$\lambda_d|_{{\rm soft}}=\lambda_d|_{{\rm soft}}^1$ as is already
stated. The soft pion correction for the $F_2^{\nu N}|_{{\rm soft}}$ 
can be calculated as
\begin{equation}
F_2^{\nu N}|_{{\rm soft}}=\frac{I_{\pi}}{f_{\pi}^2}[2(F_2^{\nu N})_0
+ \frac{3}{4}g_A^2(0)(1+\langle n \rangle)F_2^{\nu N}- 12xg_A(0)(g_1^{ep} - g_1^{en})],
\end{equation}
where the suffix 0 in the expression $(F_2^{\nu N})|_0$
means the structure function in the $SU(3)$ model as in Eq.(3).
Thus, using Eqs.(1),(4),(5) and (6) we obtain
\begin{eqnarray}
x\lambda_u|_{{\rm soft}} = \frac{I_{\pi}}{f_{\pi}^2}[\frac{5x}{12}(d_v + u_v)
+ \frac{5x}{6}(\lambda_u + \lambda_d) +\frac{x}{8}g_A^2(0)(1+3\langle n \rangle)u_v
+\frac{x}{4}g_A^2(0)d_v \\\nonumber
-\frac{x}{6}g_A(0)(\triangle u_v - \triangle d_v)
+\frac{1}{4}xg_A^2(0)(1+3\langle n \rangle)\lambda_u + \frac{1}{2}xg_A^2(0)\lambda_d] ,
\end{eqnarray}
\begin{eqnarray}
x\lambda_d|_{{\rm soft}} = \frac{I_{\pi}}{f_{\pi}^2}[\frac{5x}{12}(d_v + u_v)
+ \frac{5x}{6}(\lambda_u + \lambda_d) +\frac{x}{8}g_A^2(0)(1+3\langle n \rangle)d_v
+\frac{x}{4}g_A^2(0)u_v \\\nonumber
+\frac{5x}{6}g_A(0)(\triangle u_v - \triangle d_v)
+\frac{1}{4}xg_A^2(0)(1+3\langle n \rangle)\lambda_d + \frac{1}{2}xg_A^2(0)\lambda_u] .
\end{eqnarray}\\

Let us now consider the phase space factor $I_{\pi}$.
We assume the soft pion satisfies the following two conditions:
\begin{description}
\item[(1)]The transverse momentum satisfies $|\vec{k}^{\bot}|\leq bm_{\pi}$.
\item[(2)]Feynman scaling variable $x_{F}=2k^3/\sqrt{s}$ satisfies
$|x_{F}|\leq c$, where $s=(p+q)^2$.
\end{description}
Then, $I_{\pi}$ at high energy can be calculated explicitly as
\begin{eqnarray}
I_{\pi}&=&\frac{1}{16\pi^2}\Bigg( (b^2+1)m_{\pi}^2\log\left(
\frac{\sqrt{(1+b^2)m_{\pi}^2+\frac{c^2s}{4}}+\frac{c\sqrt{s}}{2}}
{\sqrt{(1+b^2)m_{\pi}^2+\frac{c^2s}{4}}-\frac{c\sqrt{s}}{2}}\right)\\\nonumber
&-&m_{\pi}^2\log\left( \frac{\sqrt{m_{\pi}^2+\frac{c^2s}{4}}+\frac{c\sqrt{s}}{2}}
{\sqrt{m_{\pi}^2+\frac{c^2s}{4}}-\frac{c\sqrt{s}}{2}}\right)
+c\sqrt{s}\left( \sqrt{(1+b^2)m_{\pi}^2+\frac{c^2s}{4}}-
\sqrt{m_{\pi}^2+\frac{c^2s}{4}}\right) \Bigg)   .
\end{eqnarray}
Following the previous study\cite{kore98,kore20}, we set $b=1$ and
$c=0.1$. Though a large ambiguity exists here, 
these are the parameters which explain
the pion charge asymmetry in the central region with low transverse momentum
in the experiment\cite{Bebek} fairly well\cite{kore78reso,kore82},
and gave an adequate quantity required by the Gottfried defect\cite{kore98}.
Of course these parameters should be determined more accurately.
For example, the energy dependence of the parameter $c$
should be studied by the high energy experiment such as a pion charge asymmetry
measurement in the central region with the low transverse momentum.
Now as to the mean multiplicity $\langle n \rangle$,
We take $\langle n \rangle =1+a\log(s)$. 
The parameter $a$ is fixed as 0.21 in consideration of 
the (nucleon + anti-nucleon) multiplicity in the $e^+e^-$ annihilation such that
$a\log\sqrt{s}$ with $\sqrt{s}$ replaced by the CM energy of that
reaction agrees with the multiplicity of that reaction\cite{DELPHI}.
Now we can check the magnitude of the soft pion correction for the
sea quarks if we specify the input distributions which can be used on the
right hand side of Eqs.(4),(7),and (8). The exact magnitude of the soft pion
correction greatly depends on this input. Let us first estimate the
various terms by using typical sea quark distributions 
given in Ref.\cite{GS} at $Q_0^2=4{\rm GeV}^2$.
\begin{figure}
   \centerline{\epsfbox{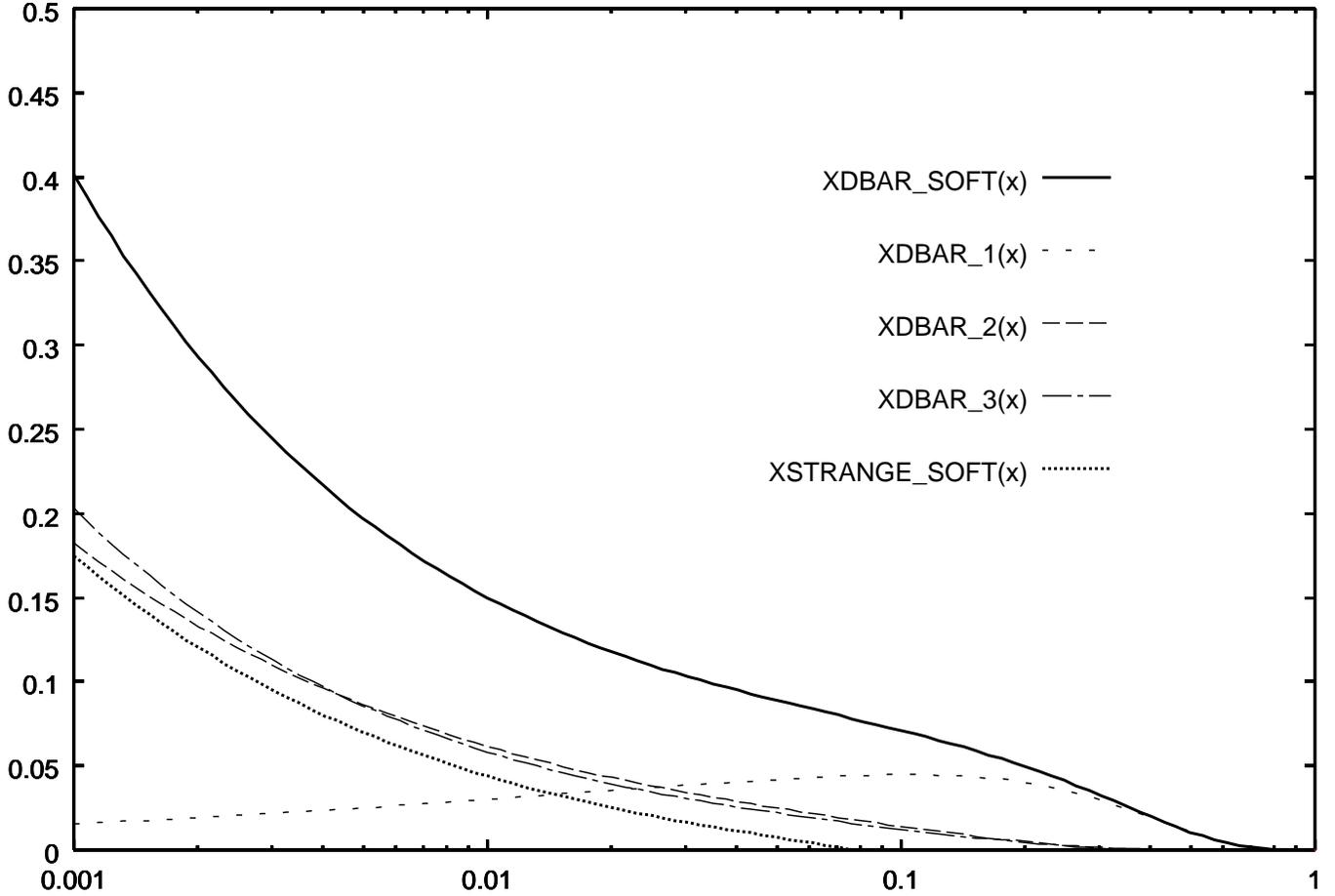}}
   \caption{A typical example of the soft pion correction for the down
sea quark distribution $x\lambda_d$ and the strange sea quark
distribution $x\lambda_s$. The soft pion correction for
$x\lambda_d$ is decomposed into three types. XDBAR\_1(x)
is the contribution from the valence quark, XDBAR\_2(x) is the one
from the sea quark from the graph (a) and (i) in Fig.2, 
and XDBAR\_3(x) is the one from the sea quark from the graph (d) in Fig.(2)}
        \label{fig:3}
\end{figure}
In FIG.3, the contribution to the $x\lambda_d|_{\rm soft}$ decomposed
into the three types are given. The XDBAR\_1(x) is the one
from the valence quarks coming from all the graphs considered in Fig.(2).
The XDBAR\_2(x) is the one from the sea quark coming from the graph $(a)$ and
$(i)$ in Fig.2. The XDBAR\_3(x) is the one from the sea quark coming from the
graph $(d)$ in Fig.2. The estimate given in Fig.3 is overestimate 
since the phenomenologically
determined quark distribution already includes the soft pion correction,
and the distributions on the right hand side of Eqs.(4),(7),and (8)
should be the ones without the soft pion correction.
However, as far as its correction is small, we can discard this fact and
study its magnitude roughly. In the region above $x \sim 0.1$, the soft
pion corrections are dominated by the terms originating from the valence quarks
and its magnitude is large in the up and the down sea quark distribution.
On the other hand, the correction for the strange sea quark distribution
is greatly suppressed in this region. This reflects the fact that
the contribution from the commutator terms is suppressed. 
Below the region $x \sim 0.1$, the corrections originating from the sea
quarks begin to become large and they take over the ones from the
valence quarks. They come both from the pion emission terms and the
commutator term. In accord with this, the correction 
for the strange sea quark distribution begins to become sizable.
We find that the correction from the soft pion for the strange sea quark
distribution is yet suppressed, but near the region $x \sim 0.01$, its magnitude
becomes about 1/3 of the correction for the up or the down sea quark one.
In the region below $x \sim 0.01$, among the above terms the
pion emission from the final nucleon(anti-nucleon) term being
proportional to $\langle n \rangle$ begins to become very
large, and, in the region below $x \sim 0.001$, its magnitude rapidly becomes
dominant one. Thus in the region below $x \sim 0.01$, by keeping the
contribution only from the sea quark, the soft pion correction
to the down sea quark distribution can be set effectively as
\begin{equation}
x\lambda_d|_{\rm soft}=\frac{I_{\pi}}{f_{\pi}^2}[\frac{5x}{3}\lambda_d+\frac{3}{4}xg_A^2(0)(1+\langle n
\rangle)\lambda_d ] ,
\end{equation}
where we set $x\lambda_u=x\lambda_d$ because,  in this small $x$ region, the
differential between the up and the down sea quark distribution 
is very small compared with their sum. Further we 
set $x\lambda_d|_{\rm soft}=x\lambda_u|_{\rm soft}$ by this same reason. 
Similarly, the strange sea qaurk distribution can be set as
\begin{equation}
x\lambda_s|_{\rm soft}=\frac{I_{\pi}}{f_{\pi}^2}[\frac{2x}{3}\lambda_d+\frac{3}{4}xg_A^2(0)(1+\langle n \rangle)\lambda_s].
\end{equation}
It should be noted that the distinction of the sea quark distributions classified
by the superscript 0 and 1 discussed in Appendix B 
does not matter since both give the same result (10).
This is because the differential between the two definitions lies in the
soft pion correction to the valence quark distribution and it is given only by the
valence quark distributions. \\
\section{The behavior of the soft pion correction in the small $x$ region}
Now the distribution on the right hand side of Eqs.(4),(7),and (8) should be
the ones without the soft pion correction as is already noted.
We can discard this fact as far as the soft pion correction
is small as in the case of the differential of the up and the 
down sea quark distributions in the modified Gottfried sum rule.
However, when it comes to the sea quark distribution itself, 
we must take into account this fact since its correction is large 
in the small $x$ region.
Here we consider this by using Eqs.(10) and (11).
In the approximation to neglect the higher order soft pion corrections,
we can express the phenomenologically determined sea qaurk distribution
$x\lambda_i|_{\rm ph}$ for $i=u,d,s$ as 
$\lambda_i|_{\rm ph}=\lambda_i|_{\rm ns}+\lambda_i|_{\rm soft}$,
where the distribution $x\lambda_i|_{\rm ns}$ is the one with no soft pion.
In this case the $x\lambda_d$ and $x\lambda_s$
on the right hand side of Eqs.(10) and (11) should be
$x\lambda_d|_{\rm ns}$ and $x\lambda_s|_{\rm ns}$ respectively.
Then it may be thought that even if $x\lambda_d|_{\rm ns}$ and $x\lambda_s|_{\rm ns}$
becomes symmetric in the very small $x$ region the soft pion
correction is asymmetric because of the difference between the first
term on the right hand side of Eq.(10) and that of Eq.(11). 
This is not the case if a certain
condition is satisfied as discussed below. 
Let us first assume $x\lambda_d|_{\rm ns}$ and $x\lambda_s|_{\rm ns}$ 
becomes symmetric somewhere in the very small $x$ region.
We call the terms proportional to $g_A^2(0)$ on the right hand
side of Eqs.(10) and (11) as symmetric terms, and the rest
as the commutator terms since they come from the graph (i) in Fig.2.
In the small $x$ region, Eq.(10) and the definition of the
phenomenologically determined distribution $\lambda_d|_{\rm ph}$
gives us the relation $x\lambda_d|_{\rm ns}=x\lambda_d|_{\rm ph}/(1+K_d)$ with 
$K_d=\frac{I_{\pi}}{f_{\pi}^2}(\frac{5}{3}+\frac{3g_A^2(0)}{4}(1+\langle
n\rangle))$.  Since $\langle n \rangle$ becomes very large as $x\to 0$,
$x\lambda_d|_{\rm ns}$ behaves as $x\lambda_d|_{\rm ns} \sim
x\lambda_d|_{\rm ph}/(\langle n \rangle I_{\pi})$ apart from
a numerical factor. Thus the commutator term in Eq.(10)
behaves as $x\lambda_d|_{\rm ph}/\langle n \rangle$,
while the rest as $x\lambda_d|_{\rm ph}
(\langle n \rangle+1)/\langle n \rangle$. Thus if the condition 
\begin{equation}
\lim_{x\to 0}\frac{x\lambda_d|_{\rm ph}}{\langle n\rangle} =0,
\end{equation}
is satisfied, the commutator term being asymmetric vanishes 
and, among the remaining symmetric term, the one which comes 
from the pion emission from the final nucleon(anti-nucleon) remains.
The important point is that this fact does not depend on the soft
pion phase space factor.
At small $Q^2 \sim 1$ GeV$^2$, the experiment at HERA
shows that the behavior of the structure function
in the small $x$ region is like the soft
pomeron. Though the nucleon multiplicity is assumed to
behave as $\log(s)$ in this paper, it can be also parameterized 
as if it behaves like $s^{0.15}$, and we cannot distinguish between 
these two cases\cite{DELPHI}. Thus the Eq.(12) have a good chance 
to be satisfied. Even if it is not satisfied, the contribution from
the commutator term becomes far smaller than that from
the pion emission terms as we go to the
smaller $x$ region. Thus the multi-soft pion effect 
from the pion emission from the nucleon(anti-nucleon)
in the final state enhances the symmetric term
and the commutator term becomes negligible. In this way,
we can understand why the soft pion correction for the sea quark
distribution becomes $SU(3)$ flavor symmetric.
Now the $SU(3)$ flavor symmetry in the limit $x\to 0$ is the necessary
condition for the mean charge sum rule which holds under the same
theoretical basis with the modified Gottfried sum rule.
It takes the form
\begin{equation}
\int_0^1dx\{\frac{2}{3}\lambda_u-\frac{1}{3}\lambda_d-\frac{1}{3}\lambda_s\}\sim
 0.2 .
\end{equation}
This sum rule is $Q^2$ independent and the
perturbative correction is negligibly small as in the modified Gottfried
sum rule. On the one hand the soft pion correction for the strange sea quark
distribution has a phenomenologically favorable property
being suppressed in the large and the medium $x$ region, but
on the other hand it can have a theoretically favorable property 
being symmetric in the very
small $x$ region. Since the sum rule is
very sensitive about the way how the symmetry of the strange sea quark distribution
is restored in the small $x$ region\cite{kore98,kore21}, 
let us study the sea quark distributions at $Q^2\sim 1$ GeV$^2$ 
quantitably by using this sum rule.
Since the sea quark distributions above $x=0.01$
can be expected to be relatively well determined phenomenologically, 
we use, for example, the distributions given by MRST\cite{MRST} in this region 
and find that it takes the value about 0.03.
Then,below $x=0.01$, we consider that $x\lambda_d|_{\rm ns}$ is determined 
through Eq.(10) as $x\lambda_d|_{\rm ns}=x\lambda_d|_{\rm ph}/(1+K_d)$,
where we use $x\lambda_d|_{\rm ph}$ as the one given by MRST.
By setting $x\lambda_s|_{\rm ns}=x\lambda_d|_{\rm ns}/2$ at $x=0.01$,
we take the interpolating function as $I(x)=1.085\exp [156.6x-5848x^2-17.65\sqrt{x}]$,
and set $\lambda_s|_{\rm ns}=I(x)\lambda_d|_{\rm ns}$ from $x=0.01$ 
up to $x=10^{-6}$, and determine the $x\lambda_s|_{\rm ph}$ by using Eq.(11) as
$x\lambda_s|_{\rm ph}=x\lambda_s|_{\rm ns}+ x\lambda_s|_{\rm soft}$. The interpolating 
function is constructed to ensure that the strange sea quark distribution
determined in this way can be continued to the one of the MRST at
$x=0.01$, and that near $x=10^{-6}$ it becomes almost symmetric 
as is seen in Fig.4. 
Then, below $x=0.01$, using this strange sea
quark distribution together with the MRST distribution for the down and the up 
sea quark distributions we find that their contribution to the sum rule is about
0.19, where the contribution below $x=10^{-6}$ is set zero by
regarding the sea quark distribution being $SU(3)$ flavor 
symmetric in this region.
Thus combining the value above $x=0.01$, the sum rule takes the value 
about 0.22. Though the extrapolation is rather arbitrary, 
we should see the behavior of the strange sea quark distribution
in the region from $x=10^{-2}$ to $10^{-6}$.  If we use the MRST sea quark
distributions even for the strange sea quark one
in the sum rule (13), their contribution to it
in this region is 0.72. Thus the sum rule is badly broken 
already in this region.

\begin{figure}
   \centerline{\epsfbox{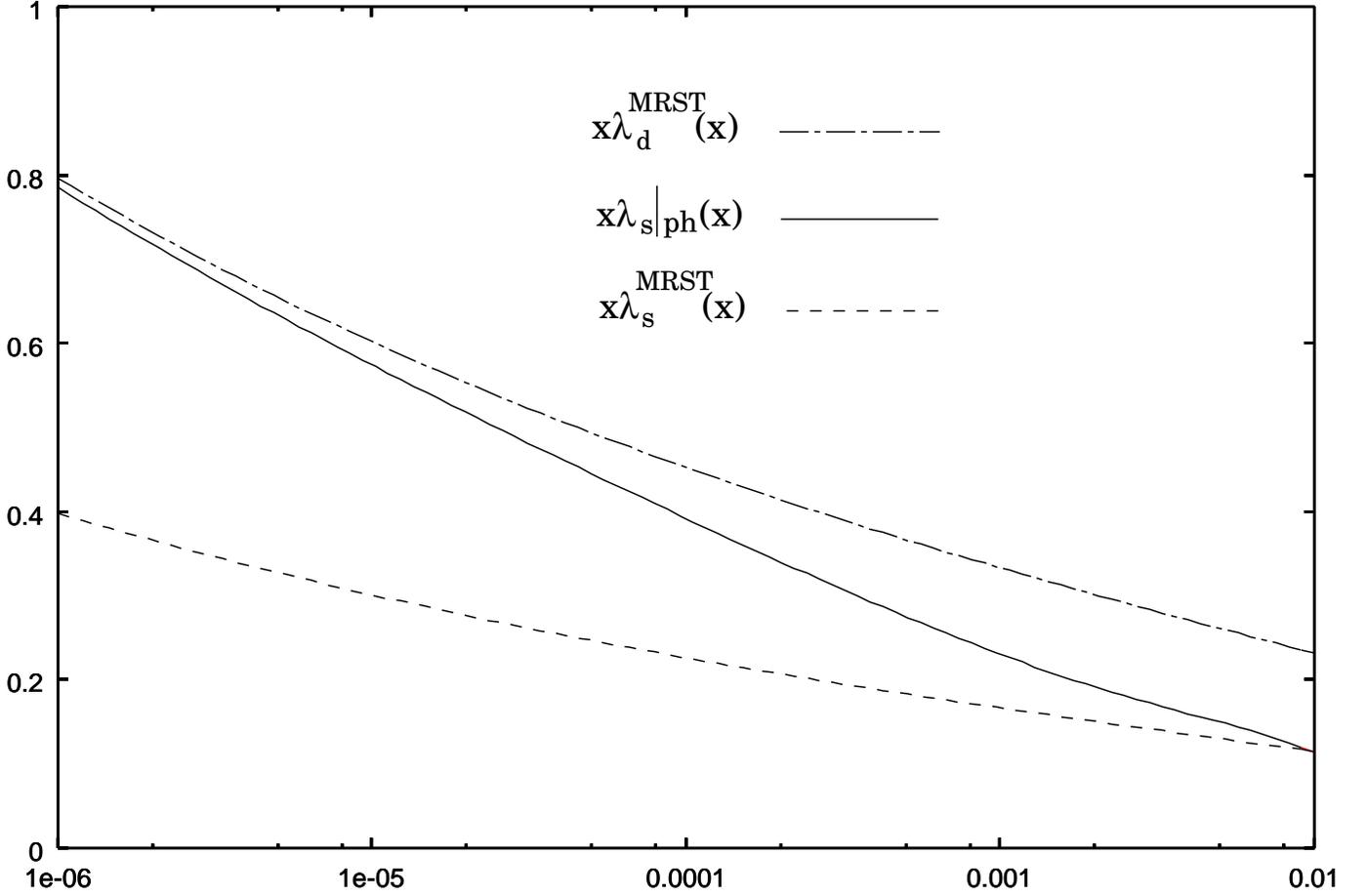}}
   \caption{The strange sea quark distribution $x\lambda_s|_{\rm ph}$ together with 
the down and the strange sea quark distribution $x\lambda_d^{MRST}$ 
and $x\lambda_s^{MRST}$ given by MRST,\cite{MRST}
where the $x\lambda_d^{MRST}$ 
is identified as the down sea quark distribution $x\lambda_d|_{\rm ph}$
in the analysis in sect.IV.}
        \label{fig:4}
\end{figure}
\section{Conclusion}
The soft pion may contribute to the experimentally measured quantity
even at high energy. If this is the case, we must either subtract
the soft pion effect from the experimental data or include it in the 
parameters of the theoretical model considered.
In this paper, we have shown that, while the soft pion correction 
for the strange sea quark distribution
is suppressed in the large and the medium $x$ region,
it becomes $SU(3)$ flavor symmetric 
in the very small $x$ region somewhere below $x\sim 0.01$. 
Based on this fact, by interpolating the asymmetric strange sea 
quark distribution to the symmetric one, 
the mean charge sum rule for the light sea quark which holds
under the same theoretical basis with the modified Gottfried sum rule
has been studied. This sum rule requires the symmetric light
sea quark distributions in the limit $x\to 0$. However, there
was no theoretical reason why the strange sea quark distribution
being suppressed in the large $x$ region became large
and flavor symmetric in the very small $x$ region. The soft pion
correction has these properties. Moreover, the sum rule is very
sensitive about the way how the symmetry of the strange sea quark
distribution is restored.
Then, by estimating the sum rule, it has been discussed that the 
large symmetry restoration of the light sea quarks originating
from the soft pion emission from the nucleon(antinucleon) in the final state
should exist in the region from $x=10^{-2}$ to $10^{-6}$ 
at $Q^2\sim 1$ GeV.
\appendix
\section{}
\label{App:alpha}
The hadronic tensor of the reaction 
$V^{\mu}(q) + N(p) \to \pi_{s}(k) + anythings(X_0)$\cite{kore78} 
can be expressed as
\begin{eqnarray}
\lefteqn{T^{\mu \nu}_{abcd}=(m_{\pi}^2-k^2)^2\int d^4xd^4yd^4z\exp [-ik\cdot (x-z)+iq\cdot y]}& \nonumber\\
\times &\langle N(p)|[T^{\dagger}(\phi_{\pi}^{a^{\prime}}(x)V^{\mu}_{b^{\prime}}(y)),T(\phi_{\pi}^{c}(z)V^{\nu}_{d}(0))]
|N(p)\rangle_c ,
\end{eqnarray}
where the spectral condition is used to express the tensor as the matrix element of 
the commutator, $a^{\prime \dagger}=a,b^{\prime \dagger}=b$, and the sum over the intermediate 
state $X_0$ is understood.
Then we define the soft pion limit of the hadronic tensor as $W^{\mu \nu}_{abcd}(p,q)$, where
we neglect the argument $k$ since the limit $k^{\mu}\to 0$ is taken. Under the exchange
$q\to -q$ and $a\leftrightarrow c, b\leftrightarrow d$, the each structure function defined by 
the hadronic tensor has a definite crossing property. 
Among the terms in $W^{\mu \nu}_{abcd}(p,q)$, 
the term coming from the graph (b) in Fig.2 is
given, for example, as
\begin{eqnarray}
\lefteqn{A^{\mu \nu}_2=\frac{-1}{4f_{\pi}^2p^+}\int d^4x\int d^4y\exp (iq\cdot y)
\delta (x^+-y^+)}\nonumber \\
\times&\{\langle N(p)|[J_a^{5+}(x),V_b^{\mu}(y)]V_d^{\nu}(0)|N^{\prime}(p)\rangle 
\langle N^{\prime}(p)|J_c^{5+}(0)|N(p)\rangle\nonumber \\
+&\langle N(p)|J_c^{5+}(0)|N^{\prime}(p)\rangle 
\langle N^{\prime}(p)|V_d^{\nu}(0)[J_a^{5+}(x),V_b^{\mu}(y)]|N(p)\rangle\} .
\end{eqnarray}
If we set $a=1+i2, c=1-i2,$ and $b=d^{\dagger}$ and take the target as the proton,
the second term in the brace on the right hand side of
Eq.(A2) is zero. While, by taking $\mu =\nu =+$
and using the current commutation relation on the null-plane,
the first term becomes the product of the currents if we replace the sum over the intermediate state $X_0$
to the complete set of the state. We can not change this product
to the commutation relation which may be obtained as the imaginary part of the retarded
product, since it changes the crossing property.  Thus we must use the method 
which can be applied directly to the light-cone dominated process such as the cut vertex formalism\cite{Mue}.
Further this example shows that the quantity which we must calculate is the nucleon matrix element
of the currents product, hence we see that this part is related to the structure function in the
total inclusive reactions. For the purpose of searching such a relation,
we can use the light-cone current algebra\cite{fg} at some $Q^2=Q_0^2$. We take this as the point where
the perturbative evolution is started. Now we encounter the symmetric
bilocal current and the antisymmetric one. The nucleon matrix element of these currents
distinguish how the quark and the antiquark contribute. Let us explain
this fact in details.\\

We define
\begin{equation}
W^{\mu \nu}_{ab}=\frac{1}{4\pi M}\int d^4x\exp [iq\cdot x]
\langle N(p)|J_a^{\mu}(x)J_b^{\nu}(0)|N(p)\rangle_c .
\end{equation}
In the light-cone limit $q^-\to \infty$, the leading term comes from the region $x^+=0$ and $x^2=0$. 
Hence we take $\vec{x}^{\bot}=0$ but $x^-$ is left arbitrary.
Using the light-cone current algebra, we see that $F_2$ is proportional 
to $\eta \widetilde{F}_{ab}(\eta )$ with $\eta =-q^2/2p\cdot q$ 
and $\widetilde{F}_{ab}(\eta )$ being defined as
\begin{equation}
F_{ab}(p\cdot x,x^2=0)=\int d\eta \exp[i\eta p\cdot x]\tilde{F}_{ab}(\eta),
\end{equation}
where $F_{ab}(p\cdot x,x^2)$ is defined as
\begin{equation}
\langle N(p)|F_{ab}^{\mu}(x|0)|N(p)\rangle_c = p^{\mu}F_{ab}(p\cdot x,x^2)+x^{\mu}\bar{F}_{ab}(p\cdot x,x^2).
\end{equation}
The bilocal currents $F_{ab}^{\mu}(x|0)$ is decomposed into the symmetric
and the antisymmetric bilocal as $F_{ab}^{\mu}(x|0)=f_{abc}S_c^{\mu}(x|0) + d_{abc}A_c^{\mu}(x|0)$,
where
\begin{eqnarray}
S_c^{\mu}(x|0)&=&\frac{1}{2}\{ :\bar{q}(x)\gamma^{\mu}\frac{\lambda_c}{2}q(0)
+\bar{q}(0)\gamma^{\mu}\frac{\lambda_c}{2}q(x):\} ,\nonumber \\
A_c^{\mu}(x|0)&=&\frac{1}{2i}\{ :\bar{q}(x)\gamma^{\mu}\frac{\lambda_c}{2}q(0)
-\bar{q}(0)\gamma^{\mu}\frac{\lambda_c}{2}q(x):\} .
\end{eqnarray}
Here the phase factor is discarded by taking the light cone
gauge$A^+=0$ for simplicity,
but the following discussion is unchanged if we do not take this gauge
and include it.
Corresponding to the decomposition of the symmetric and antisymmetric bilocals,
we define $\widetilde{F}_{ab}(\eta )=f_{abc}S_c(\eta) + d_{abc}A_c(\eta)$.
Now we expand the quark field on the null-plane $x^+=0,\vec{x}^{\bot}=0$ into the creation 
and the annihilation operator as  
\begin{equation}
q(x)=\sum_na_n\phi_n^{(+)}(x) + \sum_nb_n^{\dagger}\phi_n^{(-)}(x) ,
\end{equation}
where the sum over the subscript $n$ means the spin sum and the momentum integral collectively.
Here $(+)$ means the positive energy solution and $(-)$
the negative one.  The normal ordered product is given as
\begin{equation}
:\bar{q}(x)\gamma^+\frac{\lambda_a}{2}q(0):=
\sum_{n,m}a_n^{\dagger}a_m\bar{\phi}_n^{(+)}(x)\gamma^+\frac{\lambda_a}{2}\phi_m^{(+)}(0)
-\sum_{n,m}b_m^{\dagger}b_n\bar{\phi}_n^{(-)}(x)\gamma^+\frac{\lambda_a}{2}\phi_m^{(-)}(0).
\end{equation}
Then, when the matrix $\lambda_a/2$ is diagonal, 
we define a part contributing to the quark distribution function of the proton as
\begin{equation}
<a>f_a(x)=\frac{1}{2\pi p^+}\int_{-\infty}^{\infty}d\alpha \exp [-ix\alpha ]\langle p|
\sum_{n,m}a_n^{\dagger}a_m\bar{\phi}_n^{(+)}(y^-)\gamma^+\frac{\lambda_a}{2}
\phi_m^{(+)}(0)|p\rangle_c,
\end{equation}
and a part contributing to the antiquark one as
\begin{equation}
<\bar{a}>g_{\bar{a}}(x)=-\frac{1}{2\pi p^+}\int_{-\infty}^{\infty}d\alpha \exp [-ix\alpha ]\langle p|
\sum_{n,m}b_m^{\dagger}b_n\bar{\phi}_n^{(-)}(0)\gamma^+\frac{\lambda_a}{2}
\phi_m^{(-)}(y^-)|p\rangle_c,
\end{equation}
where $\alpha = p^+y^-$, $<a>$ is a symmetry factor originating from the flavor symmetry, 
and $<\bar{a}>=-<a>$. Here we use an abbreviated notation. For example, when $a=3$,
$\displaystyle{<a>f_a=\frac{1}{2}f_u - \frac{1}{2}f_d}$ and 
$\displaystyle{<\bar{a}>g_{\bar{a}}=-\frac{1}{2}g_{\bar{u}} + \frac{1}{2}g_{\bar{d}}}$.
Corresponding to the decomposition of the normal ordered product, we classify the matrix element of 
the bilocal current into the quark part and the antiquark part as
\begin{equation}
S_a(\alpha )=S_a^q(\alpha )+S_a^{\bar{q}}(\alpha ),\qquad A_a(\alpha )=A_a^q(\alpha )+A_a^{\bar{q}}(\alpha ).
\end{equation}
The moments becomes
\begin{eqnarray}
\lefteqn{\int_0^{\infty}dx x^{n-1}<a>f_a(x)=\frac{(n-1)!(-i)^n}{2\pi p^+}}&\nonumber \\
\times &\int_{-\infty}^{\infty}d\alpha 
\frac{1}{(\alpha - i\epsilon )^n}\langle p|
\sum_{n,m}a_n^{\dagger}a_m\bar{\phi}_n^{(+)}(y^-)\gamma^+\frac{\lambda_a}{2}\phi_m^{(+)}(0)|p\rangle_c,
\end{eqnarray}
and
\begin{eqnarray}
\lefteqn{\int_0^{\infty}dx x^{n-1}<\bar{a}>g_{\bar{a}}(x)=\frac{-(n-1)!(i)^n}{2\pi p^+}}&\nonumber \\
\times &\int_{-\infty}^{\infty}d\alpha 
\frac{1}{(\alpha + i\epsilon )^n}\langle p|
\sum_{n,m}b_m^{\dagger}b_n\bar{\phi}_n^{(-)}(y^-)\gamma^+\frac{\lambda_a}{2}\phi_m^{(-)}(0)|p\rangle_c.
\end{eqnarray}
Then, using the facts $:\bar{q}(x)\gamma^{\mu}\frac{\lambda_a}{2}q(0):=S_a^{\mu}+iA_a^{\mu}$ 
and the support property of the quark distribution function we obtain
\begin{equation}
\int_0^{1}dx x^{n-1}<a>f_a(x)=\frac{(n-1)!(-i)^n}{2\pi}
\int_{-\infty}^{\infty}d\alpha \frac{1}{(\alpha - i\epsilon )^n}[S_a^q(\alpha )+iA_a^q(\alpha )],
\end{equation}
and
\begin{equation}
\int_0^{1}dx x^{n-1}<\bar{a}>g_{\bar{a}}(x)=\frac{(n-1)!(-i)^n}{2\pi}
\int_{-\infty}^{\infty}d\alpha \frac{1}{(\alpha - i\epsilon )^n}[S_a^{\bar{q}}(\alpha )-iA_a^{\bar{q}}(\alpha )],
\end{equation}
where $S_a^{\bar{q}}(-\alpha )=S_a^{\bar{q}}(\alpha )$ 
and $A_a^{\bar{q}}(-\alpha )=-A_a^{\bar{q}}(\alpha )$ are used to obtain the
last equation. Similar equation can be obtained for the normal ordered product
$:\bar{q}(0)\gamma^{+}\frac{\lambda_a}{2}q(x):$ and we obtain,
\begin{eqnarray}
\int_{-1}^{0}dx x^{n-1}<a>f_a(x)=(-1)^{n-1}\int_0^{1}dx x^{n-1}<a>f_a(-x)\nonumber \\
=\frac{(-1)^{n-1}(n-1)!(-i)^n}{2\pi}\int_{-\infty}^{\infty}d\alpha \frac{1}{(\alpha - i\epsilon )^n}
[S_a^q(\alpha )-iA_a^q(\alpha )],
\end{eqnarray}
and
\begin{eqnarray}
\int_{-1}^{0}dx x^{n-1}<\bar{a}>g_{\bar{a}}(x)=(-1)^{n-1}\int_0^{1}dx x^{n-1}<\bar{a}>g_{\bar{a}}(-x)\nonumber \\
=\frac{(-1)^{n-1}(n-1)!(-i)^n}{2\pi}\int_{-\infty}^{\infty}d\alpha \frac{1}{(\alpha - i\epsilon )^n}
[S_a^{\bar{q}}(\alpha )+iA_a^{\bar{q}}(\alpha )].
\end{eqnarray}
Then, since we have the relation
\begin{equation}
\frac{(-1)^{(n-1)}(n-1)!}{(\alpha - i\epsilon )^n}
=\frac{d^{n-1}}{d\alpha^{n-1}}\left(\frac{1}{(\alpha - i\epsilon )}\right)
=(-1)^{(n-1)}(n-1)!P\frac{1}{\alpha^n} + i\pi\delta^{n-1}(\alpha ),
\end{equation}
at $n=even$, we obtain 
\begin{eqnarray}
<a>\int_0^1dx x^{n-1}[\{f_a(x)+f_a(-x)\} - \{g_{\bar{a}}(x)+g_{\bar{a}}(-x)]\nonumber \\
=\frac{(n-1)!(-i)^n}{\pi}P\int_{-\infty}^{\infty}d\alpha \frac{1}{\alpha^n}S_a(\alpha ),
\end{eqnarray}
\begin{eqnarray}
<a>\int_0^1dx x^{n-1}[\{f_a(x)-f_a(-x)\} + \{g_{\bar{a}}(x)-g_{\bar{a}}(-x)]\nonumber \\
=i^{n}\int_{-\infty}^{\infty}d\alpha\delta^{n-1}(\alpha )A_a(\alpha ),
\end{eqnarray}
and at $n=odd$
\begin{eqnarray}
<a>\int_0^1dx x^{n-1}[\{f_a(x)+f_a(-x)\} - \{g_{\bar{a}}(x)+g_{\bar{a}}(-x)]\nonumber \\
=i^{n-1}\int_{-\infty}^{\infty}d\alpha\delta^{n-1}(\alpha )S_a(\alpha ),
\end{eqnarray}
\begin{eqnarray}
<a>\int_0^1dx x^{n-1}[\{f_a(x)-f_a(-x)\} + \{g_{\bar{a}}(x)-g_{\bar{a}}(-x)]\nonumber \\
=\frac{(n-1)!(-i)^{n-1}}{\pi}P\int_{-\infty}^{\infty}d\alpha \frac{1}{\alpha^n}A_a(\alpha ).
\end{eqnarray}
Eqs.(A19) and (A22) correspond to the moments of the missing integers in the classical 
derivation of the moment sum rule and that they are expressed by the nonlocal quantity. 
Information of these missing parts are supplied by the cut vertex
formalism. The moment at $n=1$ stands on a particular status,
since we have another method to get information of it. This method
is very general.  It is independent of the light-cone limit and needs no
particular form of the currents. For a detail of this method and the definition of the quark
distributions in this method see Ref.\cite{kore93,kore-ichep} 
and the paper cited therein. As it is explained there, the moments of the
structure functions at $n=1$ can be decomposed into the expressions
which can be regarded as the moments of the quark distribution functions
at $n=1$. The mean charge sum rule discussed in this paper
is one example obtained by this method and it holds at any $Q^2$.
The reason why we obtain the moments 
like (A19) and (A22) lies in the fact that the current product decomposes into the 
commutator and the anticommutator. Hence we have the structure function defined by 
the current commutator and the one by the current anticommutator. They are identically the same 
in the $s$ channel but opposite in sign in the $u$ channel. Thus the crossing property 
is opposite. In terms of the quark distribution, this appears as $\int_{-1}^{1}dx x^{n-1}f_a(x)$
and $\int_{-1}^{1}dx x^{n-1}\epsilon (x)f_a(x)$, where $\epsilon (x)$ is a sign
function, and explains why we have two moments for each $n$.
The fact is important when we consider the analytical continuation
to the complex $n$ plane to obtain the anomalous dimension in the missing integer
in the classical derivation\cite{RS,kore83}.
Finally, we see that the (quark - antiquark) and the (quark + antiquark) corresponds to
the symmetric bilocal and the antisymmetric bilocal respectively. We know that the two
different combinations of the quark and the antiquark evolve differently\cite{RS,CFP}, 
hence the distinction of these two bilocals is important . 
However, under the approximation to neglect the sea quark which is equivalent to
neglect the antiquark, we need not distinguish the symmetric and antisymmetric bilocals.

\section{}
\label{App:beta}
Let us consider how the soft pion contribution enters into 
the correction to $1/3$ in the Gottfried sum. In the parton model, we have
\begin{equation}
x(\lambda_{\bar{d}} - \lambda_{\bar{u}})
= -\frac{1}{4}(F_2^{\nu p} - F_2^{\bar{\nu}p}) - \frac{3}{2}(F_2^{ep} - F_2^{en}) .
\end{equation}
Thus we can obtain the soft pion contribution to the distribution
$x(\lambda_{\bar{d}} - \lambda_{\bar{u}})$ by calculating the soft pion contribution
to the structure function $(F_2^{\nu p} - F_2^{\bar{\nu}p})$ in addition to the
structure function $(F_2^{ep} - F_2^{en})$. Then we can express
\begin{equation}
x(\lambda_{\bar{d}} - \lambda_{\bar{u}}) = x(\lambda_{\bar{d}} - \lambda_{\bar{u}})|_{\rm bare}^0
+x(\lambda_{\bar{d}} - \lambda_{\bar{u}})|_{\rm soft}^0,
\end{equation}
where the superscript 0 distinguish the difference of the definition of
the sea quark distribution as explained below and the suffix bare
means the contribution other than the soft pion one in this case.
The valence part is determined by the Adler sum rule, and we express it as 
\begin{equation}
x(u_v - d_v) = x(u_v - d_v)|_{\rm bare}+ x(u_v - d_v)|_{\rm soft},
\end{equation}
where
\begin{equation}
x(d_v - u_v)|_{\rm soft}=\frac{1}{2}(F_2^{\nu p} - F_2^{\bar{\nu}p})|_{\rm soft}.
\end{equation}
The Adler sum rule determines $\int_0^1dx\{u_v(x) - d_v(x)\}=1$. 
Now, the structure function $(F_2^{ep} -F_2^{en})$ can be expressed as 
\begin{equation}
(F_2^{ep} -F_2^{en})=\frac{x}{3}(u_v-d_v)|_{\rm bare}-\frac{2x}{3}(\lambda_{\bar{d}} - \lambda_{\bar{u}})|_{\rm bare}^0
+(F_2^{ep} -F_2^{en})|_{\rm soft}.
\end{equation}
Then, expressing the valence part by the $(u_v-d_v)$, we obtain
\begin{eqnarray}
\lefteqn{(F_2^{ep} -F_2^{en})=\frac{x}{3}(u_v-d_v)}\nonumber\\
&-\frac{x}{3}(u_v-d_v)|_{\rm soft} - \frac{2x}{3}
(\lambda_{\bar{d}} - \lambda_{\bar{u}})|_{\rm bare}^0 +(F_2^{ep} -F_2^{en})|_{\rm soft},
\end{eqnarray}
and hence
\begin{eqnarray}
\lefteqn{(F_2^{ep} -F_2^{en})=\frac{x}{3}(u_v-d_v)}\nonumber\\
&+\frac{1}{6}(F_2^{\nu p} - F_2^{\bar{\nu}p})|_{\rm soft}
-\frac{2x}{3}(\lambda_{\bar{d}} - \lambda_{\bar{u}})|_{\rm bare}^0+(F_2^{ep} -F_2^{en})|_{\rm soft}.
\end{eqnarray}
By using Eqs.(B1) and (B2), we can rewrite this expression as
\begin{equation}
(F_2^{ep} -F_2^{en})=\frac{x}{3}(u_v-d_v)-\frac{2x}{3}(\lambda_{\bar{d}} - \lambda_{\bar{u}})
\end{equation}
as it should be. However, in a phenomenological analysis, we use the valence quark distribution 
determined by the Adler sum rule from the first, 
and, in stead of Eq.(B5), we express $(F_2^{ep} - F_2^{en})$ as 
\begin{equation}
(F_2^{ep} -F_2^{en})=\frac{x}{3}(u_v-d_v)-\frac{2x}{3}(\lambda_{\bar{d}} - 
\lambda_{\bar{u}})|_{\rm bare}^1+(F_2^{ep} -F_2^{en})|_{\rm soft}.
\end{equation}
Here we discriminate the bare part of the sea quark distribution
by the superscript 1 from that specified
by the superscript 0. In this case, the soft pion contribution can be expressed simply as
\begin{equation}
x(\lambda_{\bar{d}} - \lambda_{\bar{u}})|_{\rm soft}^1=-\frac{3}{2}(F_2^{ep} -F_2^{en})|_{\rm soft}. 
\end{equation}
where $(\lambda_{\bar{d}}-\lambda_{\bar{u}})=
(\lambda_{\bar{d}}-\lambda_{\bar{u}})|_{\rm bare}^1+(\lambda_{\bar{d}}-\lambda_{\bar{u}})|_{\rm soft}^1$.
By comparing Eqs.(B5) and (B9) with use of the relation (B3), we see that
the bare part of the sea quark distribution discriminated 
by the superscript 1 includes the soft pion piece as
\begin{equation}
x(\lambda_{\bar{d}} - \lambda_{\bar{u}})|_{\rm bare}^1=x(\lambda_{\bar{d}} - \lambda_{\bar{u}})|_{\rm bare}^0 
- \frac{1}{2}x(d_v - u_v)|_{\rm soft}.
\end{equation}
Since we have $\lambda_{\bar{d}}|_{\rm bare}^0+\lambda_{\bar{d}}|_{\rm soft}^0
=\lambda_{\bar{d}}|_{\rm bare}^1+\lambda_{\bar{d}}|_{\rm soft}^1$ and similar
equation for $\lambda_{\bar{u}}$, we have
\begin{eqnarray}
x\lambda_{\bar{d}}|_{\rm soft}^1 &=& x\lambda_{\bar{d}}|_{\rm soft}^0 + \frac{x}{2}d_v|_{\rm soft},\\
x\lambda_{\bar{u}}|_{\rm soft}^1 &=& x\lambda_{\bar{u}}|_{\rm soft}^0 + \frac{x}{2}u_v|_{\rm soft},
\end{eqnarray}
where $\lambda_d=\lambda_{\bar{d}}$ and $\lambda_u=\lambda_{\bar{u}}$.
These $\lambda_u|_{\rm soft}^1$ and $\lambda_d|_{\rm soft}^1$ are expressed
as $\lambda_u|_{\rm soft}$ and $\lambda_d|_{\rm soft}$ respectively
in sects.III and IV.
Now, from Eq.(B1) we obtain the soft pion contribution classified by the 
superscript as 0 by calculating that of the neutrino reactions as
\begin{equation}
x(\lambda_{\bar{d}} - \lambda_{\bar{u}})|_{\rm soft}^0
= \frac{I_{\pi}}{f_{\pi}^2}[\frac{x}{2}(u_v -d_v)-g_A(0)x(-\triangle u_v +2\triangle d_v)] ,
\end{equation}
where we use the relation
\begin{eqnarray}
\lefteqn{(F_2^{ep} - F_2^{en})|_{\rm soft}}\nonumber \\
 &=& \frac{I_{\pi}}{4f_{\pi}^2}[g_A^2(0)(F_2^{ep} - F_2^{en})(3<n> -1)
-16xg_A(0)(g_1^{ep} - g_1^{en})] .
\end{eqnarray}
Here we assume the symmetry for unpolarized sea quark distribution 
and neglect the polarized sea quark distribution
by the same reason as explained in the text. The correction to $1/3$ depends largely on
the bare part of the sea quark distribution, and it may be possible to expect the part defined by the 
superscript 0 is not so large.  By assuming the bare part 
$x(\lambda_{\bar{d}} - \lambda_{\bar{u}})|_{\rm bare}^0$ is zero, the numerical integration 
of the $(\lambda_{\bar{d}} - \lambda_{\bar{u}})|_{\rm soft}^0$  
from $x=0.0001$ to $x=1$ gives us the value 0.11 by using the distribution 
by MRS and GS\cite{GS} at $Q_0^2=4$GeV$^2$ and the parameters $a=0.2,b=1,c=0.1$ .
Since the integral converges well in the small $x$ region, only by the soft pion contribution
we can explain the NMC deficit.
Now the contribution to the NMC deficit from the low energy region has been investigated 
extensively by the mesonic models\cite{kumano}. They have more or less related to the spontaneous chiral
symmetry breakings, and hence will be related to the soft pions in some sense. 
While the soft pion studied here contributes also at low energy. 
Some of it should be effectively taken into account in the low energy models.  From this point
of view, it is natural to consider that the sum of the soft pion contribution to the valence quark 
distribution and the bare part $x(\lambda_{\bar{d}} -
\lambda_{\bar{u}})|_{\rm bare}^0$, which is given in Eq.(B11)
as $x(\lambda_{\bar{d}} - \lambda_{\bar{u}})|_{\rm bare}^1$, is the quantity given by 
the low energy models. In this sense the additional contribution which is not included in the
low energy models is given by the 
$x(\lambda_{\bar{d}} - \lambda_{\bar{u}})|_{\rm soft}^1=-\frac{3}{2}(F_2^{ep} -F_2^{en})|_{\rm soft}$. 
The contribution of this part to the NMC deficit is about 0.03 for the parameter
$a=0.2,b=1,c=0.1$\cite{kore20}.

\end{document}